\documentstyle[11pt]{article}

\begin{document}
\newtheorem{thm}{Theorem}[section]
\newtheorem{prop}[thm]{Proposition}
\newtheorem{lemma}[thm]{Lemma}
\newtheorem{defin}[thm]{Definition}

 %
 %
 %
 %
\def\g{\gamma}
\def\m{\mu}
\def\ms{\tilde\m}
\def\l{\lambda}
\def\e{\epsilon}
\def\r{\rho}
\def\vo{\sqrt{1 + w^2 + F/t^2}}
\def\vos{\sqrt{1 + w^2 + F/s^2}}
\renewcommand{\phi}{\varphi}

\def\open#1{\setbox0=\hbox{$#1$}
\baselineskip = 0pt
\vbox{\hbox{\hspace*{0.4 \wd0}\tiny $\circ$}\hbox{$#1$}}
\baselineskip = 10pt\!}

\def\opens#1{\setbox1=\hbox{${\scriptstyle #1}$}
\baselineskip = 0pt
\vbox{\hbox{\hspace*{0.4 \wd1} $\kern-0.35em {\scriptscriptstyle \circ}$}
\hbox{${\scriptstyle #1}$}}
\baselineskip = 10pt\!}

\def\fn{\open{f}}

\def\ln{\open{\l}}
\def\mn{\open{\m}}

\def\lns{\opens{\l}}
\def\mns{\opens{\m}}

\def\be{\begin{equation}}
\def\ee{\end{equation}}
\def\bea{\begin{eqnarray}}
\def\eea{\end{eqnarray}}
\def\beas{\begin{eqnarray*}}
\def\eeas{\end{eqnarray*}}

\newcommand{\prf}{\noindent
         {\em Proof :\ }}
\newcommand{\prfe}{\hspace*{\fill} $\Box$

\smallskip \noindent}

\def\dt{\partial_t}
\def\dr{\partial_r}
\def\dw{ \partial_w}

\def\R{{\rm I\kern-.1567em R}}
\newcommand\N{{\rm I\kern-.1567em N}}
\def\supp{\mbox{\rm supp}}

\def\n#1{\left| #1 \right|}
\def\nn#1{\Vert #1 \Vert}

\title{Cosmological solutions of the Vlasov-Einstein system with spherical,
plane, and hyperbolic symmetry}
\author{Gerhard Rein\\
        Mathematisches Institut der Universit\"at M\"unchen\\
        Theresienstr.\ 39, 80333 M\"unchen, Germany
        }
\date{}
\maketitle
\begin{abstract}
The Vlasov-Einstein system describes a self-gravitating,
collisionless gas within the framework of general relativity.
We investigate the initial value problem in a cosmological setting
with spherical, plane, or hyperbolic symmetry and prove that
for small initial data solutions exist up to a spacetime singularity
which is a curvature
and a crushing singularity. An important tool in the analysis is
a local existence result with a continuation criterion saying that
solutions can be extended as long as the momenta in the support of
the phase-space distribution of the matter remain bounded.

\end{abstract}
 %
 %
 %
 %
\section{Introduction}
\setcounter{equation}{0}
When describing the evolution of self-gravitating matter fields
within the context of general relativity, the choice of the matter
model is crucial. One can, for example, describe the matter as a perfect
fluid, as dust, or as a collisionless gas. In the latter case the matter
is represented by a number density $f$ on phase-space, i.\ e., on the
tangent bundle $TM$ of the spacetime manifold $M$. The phase-space density
$f$ satisfies a continuity equation, the so-called Vlasov equation, which
says that $f$ is constant along the geodesics of the spacetime metric.
Taking the energy-momentum tensor $T^{\alpha \beta}$ generated by $f$
as the source term in Einstein's field equations, one obtains
the Vlasov-Einstein system for a self-gravitating, collisionless gas:
\[
p^\alpha \partial_{x^\alpha} f - \Gamma^\alpha_{\beta \gamma}
p^\beta p^\gamma  \partial_{p^\alpha} f = 0,
\]
\[
G^{\alpha \beta} = 8 \pi T^{\alpha \beta},
\]
\[
T^{\alpha \beta}
= \int p^\alpha p^\beta f \,\n{g}^{1/2} \,\frac{d^4 p}{m}\;,
\]
where $\Gamma^\alpha_{\beta \gamma}$ denote the Christoffel symbols
of the spacetime metric $g_{\alpha \beta}$, $\n{g}$ denotes its determinant,
$G^{\alpha \beta}$ the Einstein tensor, $x^\alpha$ are coordinates on $M$,
$p^\alpha$ the corresponding coordinates on the tangent
space, greek indices always run from 0 to 3, and
\[
m=\n{g_{\alpha \beta} p^\alpha p^\beta }^{1/2}
\]
is the rest mass of a particle at the corresponding phase-space point.

In \cite{RR} this system was investigated in the asymptotically flat,
spherically symmetric case, i.\ e., with a metric of the form
\[
ds^2 = - e^{2\m} dt^2 +
e^{2\l} dr^2 + r^2( d\theta^2 + \sin^2\theta d\phi^2),
\]
where $(t,r,\theta,\phi)$ are the usual Schwarzschild coordinates
and $\l$ and $\m$ depend only on $t$ and $r$ and vanish at
$r=\infty$. The main result was that small initial data lead
to global, geodesically complete solutions, a result which
has no analogue for perfect fluids or dust. Indeed, Christodoulou
has shown  that in the gravitational collapse of a dust
cloud naked singularities can develop even for small data \cite{Chr}.
Therefore, the Vlasov model seems to be particularly suited to describe
the behaviour of matter
in general relativity. This is further substantiated by the
fact that for the Vlasov-Poisson system, which is the Newtonian analogue
of the Vlasov-Einstein system, there are global
existence results both in the case of an isolated system, which corresponds
to the asymptotically flat case in the relativistic problem, and
in the cosmological case, cf.\ \cite{Pf,Sch} and \cite{RR2}.

In the present paper we investigate the Vlasov-Einstein
system in a cosmological setting. In order to simplify the
problem we assume that the system has a high degree of symmetry
and take the metric to be of the form
\be \label{metric}
ds^2 = - e^{2\m} dt^2 +
e^{2\l} dr^2 + t^2( d\theta^2 + \sin_\e^2\theta d\phi^2),
\ee
where
\[
\sin_\e \theta = \left\{ \begin{array}{ccl}
\sin \theta&\ \mbox{for}\ &\e = 1,\\
1&\ \mbox{for}\ &\e = 0,\\
\sinh \theta&\ \mbox{for}\ &\e =-1 ,
\end{array}
\right.
\]
$t>0$ denotes a timelike coordinate, $r \in [0,1]$, and
the functions $\l$ and $\m$ depend only on $t$ and $r$ and are periodic in
$r$.
The angular coordinates $\theta$ and $\phi$ parametrize
the surfaces of constant $t$ and $r$, which are the orbits of
the symmetry action and which are spheres in the case of spherical
symmetry $\e =1$, tori in the case of plane symmetry
$\e = 0$, and hyperbolic planes in the case of hyperbolic symmetry
$\e = -1$. They range in the domains $[0, \pi] \times [0, 2 \pi]$,
$[0, 2\pi] \times [0, 2 \pi]$, or $[0,\infty[ \times [0,2\pi]$ respectively.
It should be pointed out that the coordinates $(t,r,\theta,\phi)$
will in general not cover the whole spacetime manifold, but they do
cover a neighborhood of the singularity at $t=0$ which will allow
us to investigate the nature of this singularity.
One way to think of the above metric is to consider the Schwarzschild
metric
\[
ds^2 = - \left(1 - \frac{2 M}{r}\right) dt^2 +
\left(1 - \frac{2 M}{r}\right)^{-1} dr^2 + r^2( d\theta^2 + \sin^2\theta
d\phi^2).
\]
If one passes through the event horizon at $r=2 M$ then the $(0,0)$- and
$(1,1)$-components of the metric change sign so that the Schwarzschild radius
$r$ becomes the timelike coordinate. If one interchanges the notation for
$t$ and $r$ and compactifies the hypersurfaces of constant $t$ by making
the components of the metric periodic in $r$ one obtains a metric of the
type (\ref{metric}) with $\epsilon = 1$.

As in \cite{RR} we
restrict ourselves to the case where all particles have the same rest
mass, normalized to 1, and move forward in time, i.\ e., $f$
is supported on the submanifold
\[
PM := \left\{ g_{\alpha \beta} p^\alpha p^\beta = -1,\ p^0 >0 \right\}
\]
of the tangent bundle, which is invariant under the geodesic flow.
Due to the symmetry the distribution function $f$ can be written as
a function of $t$, $r$, $w:= e^\l p^1$, and
$F:= t^4 (p^2)^2 + t^4 \sin_\e^2 \theta (p^3)^2$.
After calculating the Vlasov equation in these variables and the non-trivial
components of the Einstein tensor
and the energy-momentum tensor and denoting by $\dot{\phantom{\l}}$ and
$\phantom{\l}'$ the derivatives of the metric components with respect
to $t$ or $r$ respectively, the complete Vlasov-Einstein system
reads as follows:
\be \label{v}
\dt f +\frac{ e^{\m - \l} w}{\vo} \dr f -
\left( \dot \l w + e^{\m - \l} \m' \vo \right) \dw f =0,
\ee
\bea
e^{-2\m} (2 t \dot \l + 1) + \e
&=&
8\pi t^2 \r , \label{f1}\\
e^{-2\m} (2 t \dot \m - 1) - \e
&=&
8\pi t^2 p, \label{f2}
\eea
\be
\m' =
- 4 \pi t e^{\m + \l} j, \label{f3}
\ee
\be
e^{- 2 \l} \left(\m'' + \m' (\m' - \l')\right)
- e^{-2\m}\left(\ddot \l + (\dot\l + 1/t)(\dot \l - \dot \m)\right)
=
8 \pi q, \label{f4}
\ee
where
\bea
\r(t,r)
&:=&
T^0_0 (t,r) = \frac{\pi}{t^2} \int_{-\infty}^\infty \int_0^\infty
\vo f(t,r,w,F)\,dF\,dw ,\label{r}\\
p(t,r)
&:=&
T^1_1 (t,r) = \frac{\pi}{t^2} \int_{-\infty}^\infty \int_0^\infty
\frac{w^2}{\vo} f(t,r,w,F)\,dF\,dw, \label{p}\\
j(t,r)
&:=&
- e^{\l - \m} T^1_0 (t,r) = \frac{\pi}{t^2}
\int_{-\infty}^\infty \int_0^\infty
w f(t,r,w,F)\,dF\,dw, \label{j}\\
q(t,r)
&:=&
T^2_2 (t,r) = \frac{\pi}{2t^4} \int_{-\infty}^\infty \int_0^\infty
\frac{F}{\vo} f(t,r,w,F)\,dF\,dw;\ \  \label{q}
\eea
the  $(3,3)$-component of the field equations which is also non-trivial
coincides with the $(2,2)$-component due to the symmetry. Note
that on $PM$ we can express $p^0$ by the other coordinates
and in the above new variables obtain
\[
p^0 = e^{-\m} \vo.
\]
Note also that $F$ is a conserved quantity of
the geodesic flow, the modulus of the angular momentum of particles,
and thus there is no $F$-derivative in the Vlasov equation.
Furthermore, the latter equation does not depend on $\e$.

We are going to study the initial value problem corresponding
to this system and prescribe initial data at time t=1,
\[
f(1,r,w,F) = \fn (r,w,F),\ \l (1,r) = \ln (r),\ \m (1,r) = \mn (r) .
\]
Our main result is that the solutions to this initial value problem
exist on the time interval $]0,1]$ provided the data satisfy a certain
smallness assumption, and we prove that the singularity at $t=0$ is
not just a coordinate singularity, but a ``real'' spacetime singularity,
a curvature and a crushing singularity.

The paper proceeds as follows: In the next section we extract a certain
subsystem from the full Vlasov-Einstein system (\ref{v})--(\ref{f4}) and
show that this subsystem is equivalent to the full system.
In Section~3 we prove a local-in-time existence and uniqueness result
for classical solutions, together with
a continuation criterion which says that when going backward in time,
i.\ e., towards the singularity a solution can be extended
as long as the support of $f$ remains bounded with respect to $w$.
The main difficulty here is to show that a solution cannot break
down due to a blow-up of a derivative of $f$ (or one of its moments),
that is to say, there is no formation of shocks. This continuation criterion
is used in Section~4 to prove that solutions exist on $]0,1]$ for
sufficiently small initial data, the support of $f$ with respect to $w$
is shown to decay like $t^c$ for some $c>0$ as $t\to 0$.
The structure of the singularity at $t=0$ is analyzed in Section~5.
In the last section
we briefly investigate the behaviour of the solutions for $t>1$.
In order to extend a solution forward in time one
needs to bound the support of $f$ with respect to $w$ and the metric
component $e^{2 \m}$. A bound on the former quantity can be established
regardless of the size of the initial data, but for $\e = 1$ it can be shown
that $e^{2 \m}$ blows up in finite coordinate time.

As to the physical relevance of the situation studied here
our point of view is that some results in general
relativity are intended to describe concrete, real-world phenomena
while others are intended to elucidate general features of the theory
like for example the structure of possible singularities. The present
paper belongs to the second category.

An approach which may generalize to other situations
more easily than the present one but gives less information on the
structure of the singularity and on questions of existence of
solutions is taken
in \cite{Ren}, where the system with spherical and plane symmetry is
analyzed using constant mean curvature slicing. As is shown in \cite{Ren2}
homogeneous solutions, i.~e., solutions which are independent of $r$ have a
curvature singularity provided $f$ is not identically zero.

To conclude this introduction we mention some further results on the
asymptotically flat, spherically symmetric Vlasov-Einstein system:
In \cite{RR3} it is shown that solutions of this system converge
to solutions of the Vlasov-Poisson system in the Newtonian limit.
In \cite{RRS} it is shown that if a singularity forms the first one
has to form at the centre of symmetry. The existence of static solutions is
established in \cite{R,RR4}.

{\bf Acknowledgements:} I would like to thank A.~D.~Rendall for helpful
discussions and comments. The present investigation was completed
during a stay at the Erwin Schr\"odinger International Institute for
Mathematical Physics in Vienna.
I would like to thank the Institute and in particular
P.~Aichelburg and R.~Beig for the kind invitation.
\section{Equivalent subsystems}
\setcounter{equation}{0}
Let us first make precise the regularity properties which we require
of a solution:
\begin{defin} \label{regdef}
Let $I \subset \R^+$ be an interval.
\begin{itemize}
\item[{\rm (a)}]
$f \in C^1(I \times \R^2 \times \R_0^+)$ is regular, if
$f(t,r+1,w,F) = f(t,r,w,F)$ for $(t,r,w,F) \in I \times \R^2 \times \R_0^+$,
$f\geq 0$ ,
and $\supp f(t,r,\cdot,\cdot)$ is compact, uniformly in
$r$ and locally uniformly in $t$.
\item[{\rm (b)}]
$\r$ (or $p,\ j,\ q$) $\in C^1 (I \times \R)$ is regular, if
$\r (t,r+1) = \r (t,r)$ for $(t,r) \in I \times \R$.
\item[{\rm (c)}]
$\l\in C^1(I \times \R)$ is regular, if $\dot \l \in C^1(I \times \R)$
and $\l (t,r+1) = \l (t,r)$ for $(t,r) \in I \times \R$.
\item[{\rm (d)}]
$\m \in C^1(I \times \R)$ is regular, if $\m'\in C^1(I \times \R)$
and $\m (t,r+1) = \m (t,r)$ for $(t,r) \in I \times \R$.
\end{itemize}
\end{defin}
We identify such functions with their restrictions to the interval $[0,1]$
with respect to $r$. The fact that regularity means different things for
different objects will cause no ambiguities.

Let $(f,\l,\m)$ be a regular solution of the subsystem (\ref{v}),
(\ref{f1}), (\ref{f2}) on an interval $I$ with $1 \in I$. We want to show
that (\ref{f3}) and (\ref{f4}) hold as well.
Integrating (\ref{f2}) we obtain
\be
t e^{- 2 \m (t,r)} = e^{- 2 \mns (r)} - \e (t - 1) - 8 \pi
\int_1^t p(s,r)\, s^2\,ds, \label{mudar}
\ee
and
\[
- 2 t \m' (t,r) e^{- 2 \m (t,r)} = - 2 \mn' (r) e^{- 2 \mns (r)}
- 8 \pi \int_1^t  p'(s,r)\,s^2\, ds .
\]
{}From (\ref{v}) and integration by parts it follows that
\beas
\int_1^t p'(s,r)\, s^2\,ds &=&
\pi\int_1^t \int_{-\infty}^\infty \int_0^\infty
\frac{w^2}{\vos} \dr f(s,r,w,F)\,dF\,dw\,ds\\
&=&
\int_1^t (\dot \l - \dot \m) e^{\l - \m} j(s,r)\, s^2\,ds
- e^{\l - \m} j(s,r) s^2 \Bigr|_{s=1}^{s=t}\\
&&
- \int_1^t \m'(s,r)\bigl(\r (s,r) + p (s,r)\bigr)\, s^2\,ds \\
&&
- 2 \int_1^t \dot \l (s,r) e^{\l - \m} j (s,r)\, s^2\,ds.
\eeas
Adding (\ref{f1}) and (\ref{f2}) yields
\be
\dot \l + \dot \m = 4 \pi t e^{2 \m} (\r + p) , \label{lpm}
\ee
and if we assume that the constraint equation (\ref{f3}) holds at time $t=1$
then these identities imply
\[
t e^{-2 \m} \left( \m' + 4 \pi t e^{\l + \m} j \right)
= - 4 \pi \int_1^t (\r + p)
\left( \m' + 4 \pi s e^{\l + \m} j \right)\, s^2\,ds
\]
so that
\[
\m' + 4 \pi t e^{\l + \m} j = 0
\]
on $I$, i.~e., (\ref{f3}) holds for all $t \in I$.
The latter equation can be differentiated with respect to $r$ to yield
\[
\m'' = (\l' + \m') \m' - 4 \pi t e^{\l + \m} j'.
\]
{}From (\ref{v}) we obtain by integration by parts the identity
\beas
j' (t,r)
&=&
\frac{\pi}{t^2} e^{\l - \m} \int_{-\infty}^\infty \int_0^\infty
\biggl[ - \vo \dt f \\
&&
\qquad + \left( \dot \l w \vo + e^{\m -\l} \m'
\bigl(1 + w^2 + F/t^2\bigr) \right) \dw f \biggr] dF\,dw \\
&=&
- \frac{\pi}{t^2} e^{\l - \m} \int_{-\infty}^\infty\int_0^\infty \vo \dt f
\,dF\,dw
- e^{\l - \m} \dot \l (\r + p) - 2 \m' j.
\eeas
Equation (\ref{f1}) can be rewritten in the form
\be \label{lddar}
\dot \l = 4 \pi t e^{2 \m} \r - \frac{1 + \e e^{2 \m}}{2 t}.
\ee
Since
\beas
\dot \r (t,r)
&=&
-\frac{2\r (t,r)}{t} -
\frac{\pi}{t^2} \int_{-\infty}^\infty\int_0^\infty \frac{F/t^3}{\vo} f
\, dF\,dw \\
&&
+   \frac{\pi}{t^2} \int_{-\infty}^\infty\int_0^\infty \vo \dt f \,dF\,dw \\
&=&
-\frac{2\r (t,r)}{t} - \frac{2 q (t,r)}{t}
+   \frac{\pi}{t^2} \int_{-\infty}^\infty\int_0^\infty \vo \dt f
\, dF\,dw,
\eeas
differentiating (\ref{lddar}) with respect to $t$ yields
\beas
\ddot \l &=&
- 4 \pi e^{2 \m} \r + 2 \dot \l \dot \m + \frac{\dot \m}{t}
- 8 \pi e^{2 \m} q \\
&& + \frac{4 \pi^2}{t} e^{2 \m} \int \vo \dt f\, dF\, dw
+ \frac{1 + \e e^{2 \m}}{2 t^2}.
\eeas
Combining these identities implies the remaining field equation (\ref{f4}).
Thus we have established the following result:
\begin{prop} \label{red1}
Let $(f,\l,\m)$ be a regular solution of $(\ref{v})$, $(\ref{f1})$,
$(\ref{f2})$ on some time interval $I \subset \R^+$ with $1 \in I$,
and let the initial data satisfy $(\ref{f3})$ for $t=1$.
Then $(\ref{f3})$ and $(\ref{f4})$ hold for all $t \in I$.
\end{prop}
When proving local existence it now suffices to consider
the subsystem  (\ref{v}), (\ref{f1}), (\ref{f2}).
However, it  would then become technically very unpleasant to control
$\m'$. It is more convenient to consider an auxiliary system, which consists
of the modified Vlasov equation
\be \label{vs}
\dt f +\frac{ e^{\m - \l} w}{\vo} \dr f -
\left( \dot \l w + e^{\m - \l} \ms \vo \right) \dw f =0,
\ee
together with (\ref{f1}), (\ref{f2}), and
\be
\ms =
- 4 \pi t e^{\l + \m} j. \label{f3s}
\ee
Assume that we have a regular solution $(f,\l,\m,\ms)$ of this system, where
regularity for $\ms$ means that this function has the same properties
as $\m'$ for $\m$ regular. We want to show that $\ms$ is nothing else
than $\m'$ so that by Proposition~\ref{red1} $(f,\l,\m)$ solves
the full system. As above,
\[
t \m'(t,r) e^{-2\m} = \mn'(r) e^{-2\mns} + 4 \pi \int_1^t p'(s,r)\, s^2 ds
\]
and
\[
\int_1^t p' s^2 ds = - \int_1^t (\dot \l + \dot \m) e^{\l -\m} j s^2 ds
- e^{\l - \m} j s^2 \Bigr|_{s=1}^{s=t}
- \int_1^t \ms (\r + p) s^2 ds .
\]
Using (\ref{lpm}) and (\ref{f3s}) we obtain
\[
\m' t e^{-2 \m} = \mn' e^{-2\mns} - 4 \pi t^2 e^{\l -\m} j +
4 \pi e^{\lns -\mns}
\open{\jmath}
\]
so that (\ref{f3}) holds for all $t \in I$ if it holds for $t=1$.
\begin{prop} \label{red2}
Let $(f,\l,\m,\ms)$ be a regular solution of $(\ref{vs})$, $(\ref{f1})$,
$(\ref{f2})$, $(\ref{f3s})$. Then $(f,\l,\m)$ solves
$(\ref{v})$--$(\ref{f4})$.
\end{prop}
We conclude this section with a result which reflects the
conservation of the number of particles in our system and is an immediate
consequence of the Vlasov equation.
\begin{prop} \label{cons}
Assume that $f$ is regular and satisfies $(\ref{v})$ with regular
coefficients $\l$ and $\m$. Then
\[
\dt \left( e^\l \int_{-\infty}^\infty \int_0^\infty f \, dF\, dw \right)
+ \dr \left(e^\m \int_{-\infty}^\infty \int_0^\infty
\frac{w}{\vo} f \, dF\, dw \right) =0
\]
and
\[
\int_0^1 \int_{-\infty}^\infty \int_0^\infty e^{\l (t,r)} \,f(t,r,w,F) \, dF\,
dw\,dr
\]
is conserved.
\end{prop}
\section{Local existence and continuation of solutions}
\setcounter{equation}{0}
In this section we prove the following local existence and uniqueness
result with the continuation criterion described in the introduction:
\begin{thm} \label{locex}
Let $\fn \in C^1(\R^2 \times \R_0^+)$ with
$\fn (r+1 ,w,F)= \fn (r,w,F)$ for $(r,w,F) \in \R^2 \times \R_0^+$,
$\fn \geq 0$,
and
\beas
w_0
&:=&
\sup \left\{ \n{w} \mid (r,w,F) \in \supp \fn \right \} < \infty,\\
F_0
&:=&
\sup \left\{ F \mid (r,w,F) \in \supp \fn \right \} < \infty .
\eeas
Let $\ln,\mn \in C^1 (\R)$
with $\ln (r) = \ln (r+1),\ \mn(r+1) = \mn(r)$ for $r\in \R$ and
\[
\mn'(r) = - 4 \pi e^{\lns + \mns} \open{\jmath}(r),\ r \in \R.
\]
In the case of hyperbolic symmetry $\e = -1$ assume in addition that
$\mn (r) <0$ for $r \in \R$.
Then there exists a unique, left maximal, regular solution $(f,\l,\m)$
of $(\ref{v})$--$(\ref{f4})$ with $(f,\l,\m)(1) = (\fn,\ln,\mn)$
on a time interval $]T,1]$ with $T \in [0,1[$.
If
\[
\sup \Bigl\{ \n{w} \mid (t,r,w,F) \in \supp f \Bigr\} < \infty
\]
then $T=0$.
\end{thm}
\noindent
{\bf Remark:}
To motivate the restriction on $\mn$ in the case $\e=-1$ consider the
following ``pseudo-Schwarzschild'' solution
\[
ds^2 = - \left(1 + \frac{2 M}{t}\right)^{-1} dt^2 +
\left(1 + \frac{2 M}{t}\right) dr^2 + t^2( d\theta^2 + \sinh^2\theta d\phi^2)
\]
with $M \in \R$ which is a vacuum solution of our system. The restriction
$\mn < 0$ is equivalent to $M>0$ so that this case is contained in the
present investigation. For $M=0$ the spacetime is flat, and for $M<0$
it has a coordinate-singularity at $t= -2 M$, but the spacetime can be
extended through this singularity, something we want to exclude from our
investigation.

\noindent
{\em Proof of Theorem~\ref{locex}:}
Define
\[
\open{\ms} := \mn',
\]
and consider the auxiliary system (\ref{vs}), (\ref{f1}), (\ref{f2}),
(\ref{f3s}). We construct a sequence of iterative solutions in the following
way:

\noindent
{\em Iterative scheme:} Let $\l_0 (t,r) := \ln (r),\ \m_0(t,r) := \mn (r),\
\ms_0 (t,r) := \open{\ms}(r)$ for $t \in ]0,1],\ r \in \R$.
If $\l_{n-1},\ \m_{n-1},\ \ms_{n-1}$ are already defined and regular on
$]0,1]\times \R$ then let
\beas
&&G_{n-1}(t,r,w,F) := \\
&& \hspace{60pt}
\left(\frac{ e^{\m_{n-1} - \l_{n-1}} w}{\vo}, -
\dot \l_{n-1} w - e^{\m_{n-1} - \l_{n-1}} \ms_{n-1} \vo \right)
\eeas
and denote by $(R_n,W_n)(s,t,r,w,F)$ the solution of the characteristic
system
\[
\frac{d}{ds}(R,W) = G_{n-1} (s,R,W,F)
\]
with initial data
\[
(R_n,W_n)(t,t,r,w,F) = (r,w),\ (t,r,w,F) \in ]0,1] \times \R^2
\times \R_0^+;
\]
note that $F$ is constant along characteristics. Define
\[
f_n (t,r,w,F) := \fn \left((R_n,W_n)(1,t,r,w,F),F\right),
\]
that is, $f_n$ is the solution of
\beas
&&
\dt f_n +\frac{ e^{\m_{n-1} - \l_{n-1}} w}{\vo} \dr f_n \\
&&
\hspace{99pt}
-
\left( \dot \l_{n-1} w + e^{\m_{n-1} - \l_{n-1}} \ms_{n-1} \vo \right) \dw
f_n =0
\eeas
with $f_n(1) =\fn$,
and define $\r_n,\ p_n,\ j_n,\ q_n$ by the integrals
(\ref{r})--(\ref{q}) with $f$ replaced by $f_n$.
Recall that the solution of (\ref{f1}) is given by (\ref{mudar}) so we define
$\m_n$ by
\be \label{mundar}
e^{-2 \m_n (t,r)} = \frac{e^{-2 \mns (r)} + \e}{t} - \e
- \frac{8 \pi}{t} \int_1^t  p_n (s,r)\, s^2\, ds ;
\ee
note that the right hand side of this equation is positive on
$]0,1] \times \R$.
It is at this point that we need
our additional assumption on the initial data  in the case $\e = -1$.
Finally define
\be \label{ldndar}
\dot \l_n (t,r) := 4 \pi t e^{2 \m_n} \r_n (t,r) -
\frac{1 + \e e^{2 \m_n}}{2t}
\ee
cf.\ (\ref{lddar}),
\[
\l_n (t,r) := \ln (r) + \int_1^t \dot \l_n (s,r) \, ds,
\]
\be \label{msndar}
\ms_n (t,r) := - 4 \pi t e^{\m_n + \l_n} j_n (t,r) .
\ee
These iterates are regular on the time interval $]0,1]$,
in particular, since $\l_{n-1},\ \m_{n-1},\ \dot \l_{n-1},\ \ms_{n-1}$ are
continuous on $]0,1]\times \R$ and periodic in $r$, these functions are
bounded on compact subintervals of $]0,1]$, uniformly in $r$, and since
$G_{n-1}$ is linearly bounded with respect to $w$ the characteristics
$R_n,\ W_n$ exist on the time interval $]0,1]$.

The proof of Theorem~\ref{locex} now consists in showing in a number of
steps that the iterates constructed above converge in a sufficiently strong
sense.

\noindent
{\em Step 1:}
As a first step we establish a uniform bound on the momenta in the support
of the distribution functions $f_n$, more precisely we want to bound
the quantities
\[
P_n (t) := \sup \Bigl \{ \n{w} \mid (r,w,F) \in \supp f_n (t) \Bigr\}
\]
uniformly in $n$. On $\supp f_n (t)$ we have
\[
\vo \leq \sqrt{1 + P_n (t)^2 + F_0/t^2} \leq \frac{1 + F_0}{t} (1 + P_n (t)),
\]
and thus
\[
\nn{\r_n(t)} \leq c \frac{(1+F_0)^2}{t^3}\nn{\fn} (1 + P_n (t))^2
\]
and
\[
\nn{p_n(t)}, \ \nn{j_n (t)} \leq c \frac{F_0}{t^2} \nn{\fn} P_n (t)^2.
\]
Throughout the paper $\nn{\cdot}$ denotes the $L^\infty$-norm on the function
space in question; we have used the fact that $\nn{f_n(t)} = \nn{\fn}$
for $n \in \N$ and $t \in ]0,1]$. The numerical constant $c$ may change from
line to line and does not
depend on $n$ or $t$ or on the initial data. In view of the continuation
criterion it is important to keep track of any dependence on the latter.
{}From (\ref{mundar}) it follows that
\[
e^{-2 \m_n (t,r)} \geq \frac{c_1}{t}
\]
where
\[
c_1 = c_1 (\mn) := \left\{
\begin{array}{lll}
\inf e^{-2 \mns} &\ \mbox{for}\ & \e =1 \ \mbox{or}\ \e =0,\\
\inf e^{-2 \mns} -1 &\ \mbox{for}\ & \e = -1.
\end{array} \right.
\]
By (\ref{ldndar}), (\ref{msndar}), and the above estimates on $\r_n$ and $j_n$
we get
\[
\left| e^{\m_n -\l_n} \ms_n (s,r) \right|
\leq 4 \pi s e^{2 \m_n} \n{j_n(s,r)} \leq c \frac{F_0}{c_1} \nn{\fn} P_n(s)^2
\]
and
\beas
\left| \dot \l_n (s,r) \right|
&\leq&
4 \pi s e^{2 \m_n} \r_n (s,r) + \frac{1 + e^{2 \m_n}}{2s} \\
&\leq&
\frac{c}{c_1} (1+F_0)^2 \nn{\fn} \frac{(1 + P_n (s))^2}{s}
+ \frac{1+1/c_1}{2s}.
\eeas
Thus
\[
\left| \dot W_{n+1}(s) \right|
\leq \frac{c_2}{s} (1 + P_n (s))^2 (1 + \n{ W_{n+1} (s)}),
\]
where
\[
c_2 = c_2 (\fn,F_0,\mn) := c\, (1+1/c_1) (1+F_0)^2 (1+\nn{\fn}).
\]
This implies that
\[
P_{n+1} (t) \leq w_0 + c_2 \int_t^1 \frac{1}{s} (1+P_n(s))^2
(1+P_{n+1}(s))\, ds.
\]
Let $z_1$ be the left maximal solution of the equation
\[
z_1 (t) = w_0 + c_2 \int_t^1 \frac{1}{s} (1 + z_1 (s))^3 ds,
\]
which exists on some interval $]T_1,1]$ with $T_1 \in [0,1[$. By induction
\[
P_n (t) \leq z_1 (t),\ t \in ]T_1,1],\ n \in \N,
\]
and all the quantities which were estimated against $P_n$ in the above
argument are bounded by certain powers of $z_1$ on $]T_1,1]$.

\noindent
{\em Step 2:}
Here we establish bounds on certain derivatives of the iterates. In particular
we need a uniform bound on the Lipschitz-constant of the right hand
side $G_n$ of the characteristic system in order to prove convergence
in the next step. Differentiating (\ref{mundar}) and (\ref{ldndar})
with respect to $r$ one obtains the identities
\beas
\m_n'(t,r)
&=&
\frac{e^{2 \m_n}}{t} \left( \mn'(r) e^{-2 \mns} + 4 \pi \int_1^t
p_n'(s,r)\, s^2\,ds \right),\\
\dot \l_n' (t,r)
&=&
e^{2 \m_n} \left( 8 \pi t \m_n'(t,r) \r_n (t,r) + 4 \pi t \r_n'(t,r)
- \frac{\e}{t}  \m_n'(t,r) \right),\\
\l_n'(t,r) &=&
\ln'(r) + \int_1^t \dot \l_n' (s,r)\, ds .
\eeas
In the following $C_1$ denotes a continuous function on $]T_1,1]$ which depends
only on $z_1$. By Step~1,
\[
\nn{\r'_n (t)},\ \nn{p'_n (t)},\ \nn{j'_n (t)} \leq C_1(t)
\nn{\dr f_n (t)}.
\]
Define
\[
D_n (t) := \sup \Bigl\{ \nn{\dr f_n (s)}\; | t \leq s \leq 1 \Bigr\}.
\]
Then the above estimates and the formulas for the derivatives of the
metric components show that
\[
\nn{\m'_n (t)},\ \nn{\l'_n (t)},\ \nn{\dot \l'_n (t)}
\leq C_1 (t) (c_3 + D_n (t)),
\]
where $c_3:=\nn{e^{-2\mns}\mn'} +\nn{\ln'} +1$.
{}From (\ref{msndar}) it follows that
\[
e^{\m_n - \l_n} \ms_n = - 4 \pi t e^{2\m_n} j_n,
\]
and
\[
\left| \left(e^{\m_n - \l_n} \ms_n \right)'(t,r) \right|
\leq C_1 (t) \bigl( c_3 + D_n (t)\bigr).
\]
We are now in the position to estimate the derivatives of $G_n$ with
respect to $r$ and $w$:
\beas
\dr G_n (t,r,w,F)
&=&
\biggl( (\m_n -\l_n)' e^{\m_n - \l_n} \frac{w}{\vo}, \\
&&
\hspace{93pt}
- (e^{\m_n - \l_n} \ms_n)' \vo - \dot \l'_n w\biggr),\\
\dw G_n (t,r,w,F)
&=&
\biggl(  e^{\m_n - \l_n}
\frac{1 + F/t^2}{\vo^3}, \\
&&\hspace{111pt}
- e^{\m_n - \l_n} \ms_n \frac{w}{\vo} - \dot \l_n \biggr),
\eeas
and thus
\beas
\n{\dr G_n (t,r,w,F)} &\leq& C_1 (t)\bigl(c_3 + D_n (t)\bigr),\\
\n{\dw G_n (t,r,w,F)} &\leq& C_1 (t)
\eeas
for $t \in ]T_1,1]$, $r \in \R$, $F \in [0,F_0]$, and $\n{w} \leq z_1(t)$.
Differentiating the characteristic system we obtain
\beas
&&
\n{\frac{d}{ds} \dr (R_{n+1}, W_{n+1})(s,t,r,w,F)} \leq \\
&&
\hspace{128pt}
C_1 (s) \left(c_3 + D_n (s)\right)\,
\n{ \dr (R_{n+1}, W_{n+1})(s,t,r,w,F)},
\eeas
and thus for $(r,w,F) \in \supp f_{n+1} (t) \cup \supp f_n (t)$
\[
\n{\dr (R_{n+1}, W_{n+1})(1,t,r,w,F)} \leq
\exp \left( \int_t^1 C_1(s)\left(c_3 + D_n (s)\right)\, ds \right).
\]
By definition of $D_n$ this implies that
\[
D_{n+1}(t) \leq \nn{\partial_{(r,w)} \fn\,}\,
\exp \left( \int_t^1 C_1(s)\left(c_3  + D_n (s)\right)\, ds \right).
\]
Let $z_2$ be the left maximal solution of
\[
z_2 (t) = \nn{\partial_{(r,w)} \fn\,}
\exp \left( \int_t^1 C_1(s)\left(c_3 + z_2 (s)\right)\, ds \right),
\]
which exists on an interval $]T_2, 1] \subset ]T_1,1]$.
Then by induction,
\[
D_n (t) \leq z_2 (t),\ t \in ]T_2,1],\ n \in \N,
\]
and all the quantities estimated against $D_n$ above can be bounded in terms
of $z_2$ on $]T_2,1]$.

\noindent
{\em Step 3:}
Let $[\delta, 1] \subset ]T_2,1]$ be an arbitrary compact subset on which the
estimates of Steps~1 and 2 hold. We will show that on such an interval the
iterates converge uniformly.
Define
\[
\alpha_n (t) := \sup \Bigl\{\nn{f_{n+1}(\tau) - f_n (\tau)}
\;| \tau \in [t,1] \Bigr\},
\]
and let $C$ denote a constant which may depend on the functions
$z_1$ and $z_2$ introduced in the previous two steps.
Then
\[
\nn{\r_{n+1}(t) - \r_n (t)},\ \nn{p_{n+1}(t) - p_n (t)},\
\nn{j_{n+1}(t) - j_n (t)}
\leq C \alpha_n (t),
\]
and thus
\beas
\nn{\l_{n+1} (t) - \l_n (t)},\ \nn{\dot \l_{n+1} (t) - \dot \l_n (t)},&&\\
\nn{\m_{n+1} (t) - \m_n (t)},\ \nn{\ms_{n+1} (t) - \ms_n (t)}
&\leq& C \alpha_n (t).
\eeas
Therefore,
\[
\n{G_{n+1} - G_n}(s,r,w,F) \leq C \alpha_{n-1} (s) .
\]
By Step~2
\[
\n{\partial_{(r,w)} G_n (s,r,w,F)} \leq C
\]
for all $s \in [\delta,1]$, $n \in \N$, and
$(r,w,F)$ with $\n{w} \leq z_1 (s)$.
For characteristics which start in $\supp \fn$ this implies
\beas
&&
\n{\frac{d}{ds} (R,W)_{n+1} -\frac{d}{ds} (R,W)_n}(s,t,r,w,F)
\leq \\
&&
\hspace{123pt}
C\,\Bigl|(R,W)_{n+1} - (R,W)_n\Bigr| (s,t,r,w,F) + C \alpha_{n-1} (s),
\eeas
and by Gronwall's inequality
\[
\Bigl|(R,W)_{n+1} - (R,W)_n\Bigr|(1,t,r,w,F)
\leq C \int_t^1 \alpha_{n-1}(s)\, ds.
\]
If we recall how $f_n$ was defined in terms of the characteristics
this implies
\[
\alpha_n (t) \leq C \int_t^1 \alpha_{n-1} (s)\, ds,\ n \geq 1.
\]
By induction we obtain
\[
\alpha_n (t) \leq C \frac{C^n (1-t)^n}{n!} \leq \frac{C^{n+1}}{n!}
\]
for $n \in \N$ and $t \in [\delta, 1]$. This implies that $f_n$ and all the
other quantities whose differences were estimated in terms of $\alpha_n$
converge on $[\delta,1]$, uniformly with respect to all their arguments.
These quantities therefore have continuous limits, but the established
convergence is not yet strong enough to conclude the differentiability
of, say, $f:= \lim_{n\to \infty} f_n$. In order to achieve the latter
we need the following lemma:
\begin{lemma} \label{thekiller}
Let $(\l,\m,\ms)$ be regular on some interval $I \subset \R^+$, and let
$(R,W)(\cdot,t,r,w,F)$
be the solution of
\[
\dot r = \frac{ e^{\m - \l} w}{\vos} ,\
\dot w = - \dot \l (s,r) w - e^{\m - \l} \ms (s,r) \vos
\]
with $(R,W)(t,t,r,w,F) = (r,w)$ for
$(t,r,w,F) \in I \times \R^2 \times \R_0^+$.
Define
\beas
\xi (s)
&:=&
e^{(\l - \m)(s,R)}\partial R (s,t,r,w,F),\\
\eta (s)
&:=&
\partial W (s,t,r,w,F) \\
&&
\hspace{46pt} + \left( \vos e^{\l - \m} \dot \l
\right)\Bigr|_{(s,(R,W)(s,t,r,w,F))}
\partial R (s,t,r,w,F)
\eeas
for $\partial \in \{ \dr, \dw\}$. Then these quantities satisfy the
following system of differential equations
\beas
\dot \xi (s) &=& a_1 \bigl(s,R(s),W(s),F\bigr)\;\xi (s) +
a_2 \bigl(s,R(s),W(s),F\bigr)\; \eta (s),\\
\dot \eta (s) &=& (a_3 + a_5) \bigl(s,R(s),W(s))(s),F\bigr)\;\xi (s) +
a_4 \bigl(s,R(s),W(s),F\bigr)\; \eta (s),
\eeas
where
\beas
a_1 (s,r,w,F)
&:=&
\frac{w^2}{1 + w^2 + F/s^2} \dot \l - \dot \m ,\\
a_2 (s,r,w,F)
&:=&
\frac{1 + F/s^2}{(1 + w^2 + F/s^2)^{3/2}},\\
a_3 (s,r,w,F)
&:=&
 - \frac{1}{s} \vos \left( \dot \l - \dot \m
+ \frac{F/s^2}{1 + w^2 + F/s^2}\dot \l \right),\\
a_4 (s,r,w,F)
&:=&
- \frac{w}{\vos} \left( e^{\m - \l} \ms + \frac{w}{\vos} \dot \l \right),\\
a_5 (s,r,w,F)
&:=&
- \vos \,e^{2\m} \\
&& \left(e^{- 2 \l} \left(\ms' + \ms (\m' - \l')\right)
- e^{-2\m}\left(\ddot \l + (\dot\l + 1/s)(\dot \l - \dot \m)\right)
\right).
\eeas
In particular, if $\ms = \m'$ and $(\l,\m)$ solves the field equations
$(\ref{f1})$--$(\ref{f4})$ then
\[
a_5 (s,r,w,F) =  - \vos \, e^{2\m(s,r)} 8 \pi q(s,r) .
\]
\end{lemma}
The proof is only a lengthy calculation and therefore omitted. The lemma
will be used twice in the further argument: In the next step it will be
used to prove that also certain derivatives of the converging sequences
obtained in the previous step converge, thus obtaining a regular, local
solution. Then it will be used in Step~6 to show that control
on the support of the solution with respect to $w$ suffices to extend
the solution to the interval $]0,1]$.

\noindent
{\em Step~4:}
Fix $\delta \in ]T_2,1]$ and $U>0$, and consider the
system derived in Lemma~\ref{thekiller} with $(\l_n,\m_n,\ms_n)$ instead
of $(\l,\m,\ms)$, and call the corresponding coefficients $a_{n,i}$,
$i=1,\ldots,5$. By Steps~1 and 2 we have the estimates
\be \label{cest}
\n{a_{n,i} (t,r,w,F)} + \n{\partial_{(r,w)} a_{n,i} (t,r,w,F)}
\leq C
\ee
for $n\in \N$, $i=1,\ldots,4$, $0\leq F \leq F_0$, $\n{w} \leq U$, and
$t \in [\delta,1]$. The only new terms to estimate here are $\dot \m_n$
and
$\dot \m'_n$, but from (\ref{mundar}) we obtain
\[
\dot\m_n = 4 \pi t e^{2 \m_n} p_n + \frac{1 + \e e^{2 \m_n}}{2 t},
\]
and
\[
\dot \m'_n = 2 \m'_n (\dot \m_n - \frac{1}{2t}) + 4 \pi t e^{2 \m_n} p'_n,
\]
so both of these terms are bounded by Steps~1 and 2. The convergence
etablished in Step~3 shows that
\[
a_{n,i} (t,r,w,F) - a_{m,i} (t,r,w,F) \to 0,\ n,m \to \infty,\ i=1,\ldots,4,
\]
uniformly on $[\delta,1]\times \R \times [-U,U] \times [0,F_0]$. The
crucial term in the present argument is $a_{n,5}$, more precisely
the expression
\[
g_n := e^{- 2 \l_n} \left(\ms'_n + \ms_n (\m'_n - \l'_n)\right)
- e^{-2\m_n}\left(\ddot \l_n +
(\dot\l_n + 1/t)(\dot \l_n - \dot \m_n)\right).
\]
If the iterates solved the field equation (\ref{f4}) then this term would
equal $8 \pi q_n$ and would also converge. The idea how to treat
$a_{n,5}$ is to show that
$g_n - 8 \pi q_n \to 0$ for $n \to \infty$ and then use the fact that
$q_n$ converges and has uniformly bounded $r$-derivative.
Now
\[
\ms'_n = (\m'_n + \l'_n)\ms_n - 4 \pi t e^{\m_n +\l_n} j'_n
\]
and
\[
\ddot \l_n = 2 \dot \l_n \dot\m_n + \frac{\dot\m_n}{t} +
4 \pi t e^{2\m_n} \left( \dot \r_n + \frac{\r_n}{t} \right)
+ \frac{1 + \e e^{2 \m_n}}{2 t^2}.
\]
{}From the definition of $\r_n$ we obtain
\beas
\dot \r_n &=&
- \frac{2}{t} \r_n - \frac{2}{t} q_n \\
&&
- e^{\m_{n-1} - \l_{n-1}} j'_n - 2 \ms_{n-1} e^{\m_{n-1} - \l_{n-1}} j_n
- \dot \l_{n-1} (\r_n + p_n),
\eeas
where we used the Vlasov equation to express $\dt f_n$ and integrated by
parts; note that the coefficients in that equation have index $n-1$.
Inserting all this into the expression for $g_n$ yields, after cancelling
a number of terms,
\beas
g_n &=& 2 e^{-2 \l_n} \ms_n \left( \m'_n - e^{\m_{n-1}-\m_n +\l_n -\l_{n-1}}
\ms_{n-1} \right) \\
&& + 4 \pi t j'_n \left( e^{\m_{n-1}-\l_{n-1}} - e^{\m_n -\l_n} \right)\\
&& + e^{-2 \m_n} (\dot \l_n + \dot \m_n )(\dot \l_{n-1} - \dot\l_n)
+ 8 \pi q_n .
\eeas
By Steps~1, 2, and 3 it remains to show that $\m_n' - \ms_{n-1} \to 0$
for $n \to \infty$ in order to conclude that $g_n \to 8 \pi q_n$. To see the
former differentiate (\ref{mundar}) with respect to $r$
to obtain
\[
\m'_n = \frac{e^{2 \m_n}}{t} \left( -\mn' e^{-2 \mns} + 4 \pi
\int_1^t p'_n (s,r) \,s^2\,ds \right).
\]
Differentiating the defining integral of $p_n$, using the Vlasov equation
for $f_n$ to express $\dr f_n$, and integrating by parts with respect to
$w$ and $s$ results in the relation
\beas
&&
\m'_n
=
\frac{e^{2 \m_n}}{t} \left( -\mn' e^{-2 \mns} +
4 \pi e^{\lns - \mns} \open{\jmath}\right) +
t e^{\m_n-\m_{n-1} -\l_n +\l_{n-1}} \ms_n  \\
&&
+ \frac{e^{2 \m_n}}{t} \int_1^t s e^{-2 \m_n}
\left[ e^{\m_n-\m_{n-1} -\l_n +\l_{n-1}} (\dot\l_{n-1} + \dot\m_{n-1})
\ms_n - (\dot\l_n + \dot\m_n)
\ms_{n-1} \right] ds,
\eeas
and since the initial data satisfy the constraint (\ref{f3}),
$\m'_n \to \ms$ for $n \to \infty$, in particular, $\m'_n - \ms_{n-1}
\to 0$ for $n \to \infty$.

In Step~3 we have shown that among other quantities the characteristics
$(R_n,W_n)(1,t,r,w,F)$ converge. Now for any $\varepsilon>0$ there exists
$N \in \N$ such that for $n,m \geq N$, $ s \in [\delta,1]$,
$r \in \R$, $\n{w} \leq U$,
and $F \in[0,F_0]$ we have
\[
\n{a_{n,5} (s,r,w,F) - (- 8 \pi) e^{2 \m_n} \vos q_n (s,r)} \leq \varepsilon
\]
and
\[
\n{\dr q_n (s,r)} \leq C,\ \n{q_n (s,r) - q_m (s,r)} \leq \varepsilon,
\]
which together with the estimates (\ref{cest}) implies
that
\[
\n{\dot \xi_n - \dot \xi_m}(s) + \n{\dot \eta_n - \dot \eta_m}(s) \leq
C \,\varepsilon +
C \,\n{\xi_n - \xi_m}(s) + C\,\n{ \eta_n - \eta_m}(s) .
\]
This implies the convergence of $\partial_{(r,w)}(R_n,W_n)(1,t,r,w,F)$;
note that the transformation from $\partial (R,W)$ to $(\xi,\eta)$
in Lemma~\ref{thekiller} is invertible, and the coefficients
in the transformation are convergent in the present situation.
Thus the limiting characteristic $(R,W)(1,t,r,w,F)$ and therefore also
$f$ are continuously differentiable
with respect to $r$ and $w$. This in turn implies that all the moments
calculated from $f$ are differentiable with respect to $r$, the coefficients
in the limiting characteristic system are continuously differentiable
with respect to $r,w$, and $F$, and thus  $(R,W)(1,t,r,w,F)$ is also
differentiable with respect to $F$ and $t$, and the regularity of the limit
$(f,\l,\m)$ is established.

\noindent
{\em Step 5:}
The estimates on the difference of two consecutive iterates derived  in
Step~3 can also be used on the difference of two solutions $f$ and
$g$ with the same initial data. This results in the estimate
\[
\sup \Bigl\{\nn{f(\tau) - g(\tau)} \;| \tau \in [t,1] \Bigr\}
\leq C \int_t^1
\sup \Bigl\{\nn{f(\tau) - g(\tau)} \;| \tau \in [s,1] \Bigr\}\, ds
\]
on any compact subinterval of $]0,1]$ on which both solutions exist,
and thus $f=g$ there.

\noindent
{\em Step 6:}
To conclude the proof of Theorem~\ref{locex} it remains to establish
the continuation criterion.
Let $(f,\l,\m,\ms)$ be a left maximal solution of the auxiliary system
(\ref{vs}), (\ref{f1}), (\ref{f2}), (\ref{f3s}) with existence interval
$]T,1]$.
By Proposition~\ref{red1} $(f,\l,\m)$ solves (\ref{v})--(\ref{f4}).
Now assume that
\[
P^\ast := \sup \Bigl\{\n{w} | (r,w,F) \in \supp f(t),\ t \in ]T,1] \Bigr\} <
\infty .
\]
We want to show that $T=0$ so let us assume that $T>0$ and take
$t_0 \in ]T,1[$. We will show that the system has a solution
with initial data $(f(t_0),\l(t_0),\m(t_0))$ prescribed at $t=t_0$
which exists on an interval
$[t_0 -\delta, t_0]$ with $\delta >0$ independent of $t_0$. By moving
$t_0$ close enough to $T$ this would extend our initial solution
beyond $T$, a contradiction to the initial solution being left maximal.

Steps~1--5 have shown that such a solution exists at least on the left
maximal existence interval of the solutions $(z_1, z_2)$ of
\beas
z_1 (t)
&=&
W_0 + c_2 \int_t^{t_0} \frac{1}{s} (1 + z_1 (s))^3 ds,\\
z_2 (t)
&=&
\nn{\partial_{(r,w)} f (t_0)}
\exp \left(\int_t^{t_0} C_1 (s)(c_3 + z_2(s))\, ds\right),
\eeas
where
\[
W_0 := \sup \Bigl\{ \n{w} \mid (r,w,F) \in \supp f(t_0) \Bigr\},
\]
\[
c_1 = c_1 (\m (t_0)) := \left\{
\begin{array}{lll}
\inf e^{-2 \m(t_0)} &\ \mbox{for}\ & \e =1 \ \mbox{or}\ \e =0,\\
\inf e^{-2 \m(t_0)} -1 &\ \mbox{for}\ & \e = -1,
\end{array} \right.
\]
\beas
c_2
&=&
c_2 (f (t_0),F_0,\m(t_0)) := c\, (1+1/c_1) (1+F_0)^2 (1+\nn{f (t_0)}),\\
c_3
&=&
\nn{e^{-2\m (t_0)} \m' (t_0)} + \nn{\l'(t_0)} + 1 ,
\eeas
and the function $C_1$ depends on $z_1$. Now $W_0 \leq P^\ast$,
$\nn{f (t_0)} = \nn{\fn}$, $F_0$ is unchanged since $F$ is constant along
characteristics, and (\ref{mudar}) shows that $c_1(\m(t_0)) \geq c_1(\mn)$.
Thus there exists a constant $c_2^\ast >0$ such that
$c_2 (f(t_0), F_0, \m (t_0))/s \leq c^\ast_2$ for $t_0 \in ]T,1]$ and
$s \in [T/2,1]$. Let $z^\ast_1$ denote the left maximal solution of
\[
z_1^\ast (t) = P^\ast + c_2^\ast \int_t^{t_0} (1 + z_1^\ast (s))^3 ds.
\]
Next observe that the coefficients $a_1,\ldots,a_5$ in
Lemma~\ref{thekiller} are uniformly bounded $]T,1]$ along characteristics in
$\supp f$ if we let $\ms = \m'$
and use the field equation (\ref{f4}). The lemma then shows that
\[
D^\ast := \sup \Bigl\{ \nn{\partial_{(r,w)} f (t)} | T < t \leq 1 \Bigr\}
< \infty.
\]
{}From
\beas
\m'(t,r)
&=&
\frac{e^{2 \m}}{t} \left( \mn'(r) e^{-2 \mns} + 4 \pi \int_1^t
p'(s,r)\, s^2\,ds \right),\\
\dot \l' (t,r)
&=&
e^{2 \m} \left( 8 \pi t \m'(t,r) \r (t,r) + 4 \pi t \r'(t,r)
- \frac{\e}{t}  \m'(t,r) \right),\\
\l' (t,r)
&=&
\ln' (r) + \int_1^t \dot \l' (s,r)\, ds
\eeas
we obtain a uniform bound $c_3 (\m (t_0),\l (t_0)) \leq c_3^\ast$. Let
$z_2^\ast$ be the left maximal solution of
\[
z_2^\ast (t) =  D^\ast
\exp \left(\int_t^{t_0} C_1^\ast (s)\left(c_3^\ast + z_2^\ast (s)\right)\,
ds\right),
\]
where
$C_1^\ast$ depends on $z_1^\ast$ in the same way as $C_1$ depends on $z_1$.
Clearly, $z_1^\ast$ and $z_2^\ast$ exist on an interval $[t_0 -\delta, t_0]$
with $\delta>0$ independent of $t_0$. If we choose $\delta < T/2$ then
$z_1 \leq z_1^\ast$ and $z_2 \leq z_2^\ast$ by construction, in particular,
$z_1$ and $z_2$  exist on $[t_0 -\delta,t_0]$, and the proof of the
continuation criterion  and of Theorem~\ref{locex} is complete. \prfe
\section{Existence up to $t=0$}
\setcounter{equation}{0}
In this section we show that the solutions obtained in the previous
section exist on the interval $]0,1]$ provided the initial data
are sufficiently small.
\begin{thm} \label{glex}
Let $(\fn,\ln,\mn)$ be initial data as in Theorem~$\ref{locex}$,
and assume that
\[
c:= \frac{1}{2} - 10\, \pi^2 w_0 F_0 \sqrt{1 + w_0^2 + F_0}\, \nn{e^{2\mns}}
\,\nn{\fn} > 0
\]
in case $\e =0$ or $\e =1$, and
\[
c:= \frac{1}{2}\left(1 - \nn{e^{2\mns}}\right) - 10\, \pi^2 w_0 F_0 \sqrt{1 +
w_0^2 + F_0} \,\frac{\nn{e^{2\mns}}}{1 - \nn{e^{2\mns}}}
\,\nn{\fn} > 0
\]
in case $ \e = -1$---note that $\nn{e^{2\mns}} <1$ in this case by the
assumption in Theorem~$\ref{locex}$.
Then the corresponding solution exists on the interval $]0,1]$, and
\[
\n{w} \leq w_0 t^c , \ (r,w,F) \in \supp f(t),\ t \in ]0,1] .
\]
\end{thm}
\prf
Let $(\fn, \mn,\ln)$ be initial data satisfying the smallness assumption, and
define
\[
P(t) := \sup \Bigl\{ \n{w} \mid (r,w,F) \in \supp f(t) \Bigr\},
\ t \in ]T,1].
\]
The characteristic system for (\ref{v}) and the field equations
(\ref{f1}) and (\ref{f3}) imply that
\beas
\dot w &=& - \dot \l w - e^{\m -\l} \m' \vo \\
&=&
4 \pi t e^{2 \m} \left( j \vo - \rho w\right) + \frac{1 + \e e^{2\m}}{2t} w .
\eeas
Assume that $P(t) \leq w_0$ for some $t \in ]T,1]$,
which is true at least for $t=1$. Then
\beas
0 \leq \r (t,r)
&\leq &
\frac{\pi}{t^2} \int_{-w_0}^{w_0} \int_0^{F_0}
\vo f(t,r,w,F)\, dF\, dw\\
&\leq&
2 \pi w_0 F_0 \sqrt{1 + w_0^2 + F_0}\, \nn{\fn} \,t^{-3} ,
\eeas
and
\beas
j(t,r)
&\leq&
\frac{\pi}{t^2} \int_0^{P(t)} \int_0^{F_0}
w f(t,r,w,F)\, dF\, dw
\leq
\frac{\pi}{2} w_0 F_0  \nn{\fn}\, P(t)\, t^{-2} ,\\
j(t,r)
&\geq&
\frac{\pi}{t^2} \int^0_{-P(t)} \int_0^{F_0}
w f(t,r,w,F)\, dF\, dw
\geq
- \frac{\pi}{2} w_0 F_0  \nn{\fn}\, P(t) \, t^{-2} .
\eeas
Next we have the estimate
\[
e^{- 2 \m (t,r)} = \frac{e^{- 2 \mns (r)} + \e}{t} - \e - 8 \pi
\int_1^t s^2 p(s,r)\, ds
\geq
\frac{e^{- 2 \mns (r)} + \e}{t} - \e
\]
so that
\[
e^{- 2 \m (t,r)} \geq \frac{c_1}{t}
\]
where
\[
c_1 := \left\{
\begin{array}{lcl}
\inf e^{-2 \mns } &\ \mbox{for}\ &\e =0\ \mbox{or}\ \e= 1 ,\\
\inf e^{-2 \mns } -1 &\ \mbox{for}\ &\e = - 1 .
\end{array}
\right.
\]
Thus $e^{ 2 \m (t,r)} \leq c_1^{-1} t$, and
\[
\frac{1 + \e e^{2\m}}{2t} \geq \frac{1}{2t}=: \frac{c_2}{t}
\]
for $\e =0$ or $\e=1$,
\[
\frac{1 + \e e^{2\m}}{2t} \geq \frac{1}{2t}
\left( 1 - \frac{1}{1 + c_1 t^{-1}}
\right) \geq  \frac{1}{2} \frac{c_1}{1 + c_1}
\frac{1}{t}=:  \frac{c_2}{t}
\]
for $\e = -1$. Assume that
$ w(t) >0$. Then
\bea
\dot w(t)
&\geq&
\left( c_2 -8 \pi^2  w_0 F_0 \sqrt{1 + w_0^2 + F_0}\, \frac{1}{c_1}
\,\nn{\fn} \right) \frac{w(t)}{t}\nonumber  \\
&&
- 2 \pi^2 w_0 F_0 \sqrt{1 + w_0^2 + F_0}\, \frac{1}{c_1}
\,\nn{\fn}\, \frac{P(t)}{t} ,\label{wdbelow}
\eea
whereas if $ w(t) <0$ we obtain the estimate
\bea
\dot w(t)
&\leq&
\left( c_2 -8 \pi^2 w_0 F_0 \sqrt{1 + w_0^2 + F_0}\, \frac{1}{c_1}\,
\nn{\fn} \right) \frac{w(t)}{t},\nonumber\\
&&
+ 2 \pi^2 w_0 F_0 \sqrt{1 + w_0^2 + F_0}\, \frac{1}{c_1}\,
\nn{\fn}\, \frac{P(t)}{t} . \label{wdabove}
\eea
If we let $t=1$ in (\ref{wdbelow}) and (\ref{wdabove})
our smallness condition on the initial data implies that
there exists a small constant $\delta >0$ such that
$ \dot w(1) > 0$ if $w_0/(1+\delta) \leq w(1) \leq w_0$, and $ \dot w(1) <0$
if $-w_0 \leq w(1) \leq - w_0/(1+\delta)$. This implies that
$P(t) < w_0$ on some interval $]t_0,1[$ which we choose maximal with
this property. On the interval $]t_0,1]$ the estimates (\ref{wdbelow})
and (\ref{wdabove}) hold for any characteristic which runs in $\supp f$
and for which $w(t)>0$ or $w(t)<0$ respectively.

Let $t \in ]t_0,1]$ be such that $w(t) >0$ for a characteristic in $\supp f$,
and choose $t_1 >t$ maximal with $w(s) >0$ for $s \in [t,t_1[$.
Then (\ref{wdbelow}) holds on $[t,t_1[$ which by Gronwall's inequality
implies that
\beas
w(t)
&\leq&
\exp\left( c_3 \int_{t_1}^t \frac{ds}{s} \right)
\left[ w(t_1) - c_4 \int_{t_1}^t  \exp\left( - c_3 \int_{t_1}^s
\frac{d\tau}{\tau} \right) \frac{P(s)}{s} ds \right]\\
&=&
\left(t /t_1\right)^{c_3} \left[ w(t_1) + c_4 {t_1}^{c_3}
\int_t^{t_1} s^{-1-c_3} P(s)\,ds \right],
\eeas
where
\beas
c_3
&:=&c_2 -8 \pi^2 w_0 F_0 \sqrt{1 + w_0^2 + F_0} \,\frac{1}{c_1}\, \nn{\fn},\\
c_4
&:=&
2 \pi^2 w_0 F_0 \sqrt{1 + w_0^2 + F_0}\, \frac{1}{c_1}\, \nn{\fn}.
\eeas
If $t_1 =1$ then
\[
w(t) \leq t^{c_3}  \left( w_0 + c_4
\int_t^1 s^{-1-c_3} P(s)\,ds \right).
\]
If $t_1 < 1$ then $w(t_1) =0$, and
\beas
w(t)
&\leq&
 \left(t/t_1\right)^{c_3} c_4 t_1^{c_3}
\int_t^{t_1} s^{-1-c_3} P(s)\,ds\\
&\leq&
t^{c_3}  \left( w_0 + c_4
\int_t^1 s^{-1-c_3} P(s)\,ds \right).
\eeas
Consider now $t \in ]t_0,1]$ such that $w(t) <0$, and choose $t_1 > t$
maximal with $w(s)< 0$ for $s \in [t,t_1[$. Repeating the above argument,
but now using (\ref{wdabove}) instead of (\ref{wdbelow}), yields the estimate
\[
w(t)
\geq
\left(t/t_1\right)^{c_3}  \left( w(t_1) - c_4 {t_1}^{c_3}
\int_t^{t_1} s^{-1-c_3} P(s)\,ds \right),
\]
and distinguishing the cases $t_1=1$ and $t_1 <1$ as above implies
that
\[
- w(t) \leq t^{c_3}  \left( w_0 + c_4
\int_t^1 s^{-1-c_3} P(s)\,ds\right) .
\]
Therefore,
\[
P(t) \leq t^{c_3} \left( w_0 + c_4 \int_t^1 s^{-1-c_3} P(s)\, ds\right)
\]
for all $t \in ]t_0,1]$. Applying Gronwall's inequality again
yields the estimate
\[
P(t) \leq w_0 t^{c_3 - c_4},\ t \in ]t_0,1],
\]
since $z(t) :=w_0 t^{c_3 - c_4}$ is the  solution of the integral equation
\[
z(t) = t^{c_3} \left( w_0 +  c_4 \int_t^1 s^{-1-c_3} z(s)\, ds \right)
\]
which is equivalent to the initial value problem
\[
\dot z (t) = (c_3 - c_4) \frac{z(t)}{t},\ z(1) = w_0.
\]
The estimate on $P(t)$ implies in particular that $t_0=T$, i.\ e., it holds
on the whole existence interval of the solution, which by Theorem~\ref{locex}
implies $T=0$, and the proof is complete. \prfe
\section{The nature of the singularity at $t=0$}
\setcounter{equation}{0}
In this section we investigate the behaviour of solutions as $t \to 0$.
Theorem~\ref{glex} shows that there are solutions which exist
on $]0,1]$ so that we do not investigate the empty set. However,
the results which follow do for the most part not depend on the smallness
assumption which we used in the previous section.
First we show that solutions which exist
on $]0,1]$ have a curvature singularity at $t=0$, more precisely:
\begin{thm} \label{curvsin}
Let $(f,\l,\m)$ be a regular solution of $(\ref{v})$--$(\ref{f4})$
on the interval $]0,1]$. Then
\[
\left(R_{\alpha\beta\gamma\delta} R^{\alpha\beta\gamma\delta}\right) (t,r)
\geq 6\, \left(\inf e^{-2 \mns} +\e \right) \; t^{-6},\
t \in ]0,1],\ r \in \R,
\]
where $R_{\alpha\beta\gamma}^{\phantom{\alpha\beta\gamma}\delta}$
denotes the Riemann curvature tensor corresponding to the metric
given by $\l$ and $\m$.
\end{thm}
The curvature scalar considered here is sometimes
called Kretschmann scalar.

\prf
It can be shown that
\beas
R_{\alpha\beta\gamma\delta} R^{\alpha\beta\gamma\delta}
&=&
4 \left( e^{- 2 \l} \left(\m'' + \m' (\m' - \l')\right)
- e^{-2\m}\left(\ddot \l + \dot\l (\dot \l - \dot \m)\right)
\right)^2 \\
&&
+ \frac{8}{t^2} \left( e^{-4\m} \dot \l^2 + e^{-4 \m} \dot \m^2
- 2 e^{-2(\l + \m)} (\m')^2 \right) \\
&&
+ \frac{4}{t^4} \left( e^{-2 \mu} + \e \right)^2 \\
&=:&
K_1 + K_2 + K_3 .
\eeas
One way to see this is to use a computer algebra system like Maple,
another is to exploit the symmetry of the metric
and split the summation
\[
R_{\alpha\beta\gamma\delta} R^{\alpha\beta\gamma\delta}
=
R_{abcd} R^{abcd}
+
R_{ABCD} R^{ABCD}
+
4 R_{AbCd} R^{AbCd},
\]
where lower case Latin indices take the values $0$ and $1$, and upper case
Latin indices take the values $2$ and $3$.
Now
\beas
R_{abcd}
&=&
\left( e^{- 2 \l} \left(\m'' + \m' (\m' - \l')\right)
- e^{-2\m}\left(\ddot \l + \dot\l (\dot \l - \dot \m)\right)\right)
\left( g_{ac} g_{bd} - g_{bc} g_{ad} \right),\\
R_{ABCD}
&=&
t^{-2} \left(e^{-2 \m} + \e \right)
\left( g_{AC} g_{BD} - g_{BC} g_{AD} \right),\\
R_{AbCd}
&=&
t^{-1} g_{AC} \Gamma_{bd}^0 ,
\eeas
which can be seen after a lengthy calculation. Inserting these expressions
into the above summations yields the formula for the Kretschmann scalar.
The first term $K_1$ is clearly nonnegative and can be dropped.
In order to estimate $K_2$ we insert the expressions
\beas
e^{-2\m} \dot \l
&=&
4 \pi t \r - \frac{\e + e^{-2 \m}}{2t},\\
e^{-2\m} \dot \m
&=&
4 \pi t  p + \frac{\e + e^{-2 \m}}{2t},\\
e^{-\m -\l} \m'
&=&
- 4 \pi t j
\eeas
into the formula for $K_2$ and obtain
\[
\frac{t^2}{8} K_2
=
16 \pi^2 t^2 (\r^2 +p^2 - 2 j^2) - 4 \pi t (\r -p) \frac{\e +e^{-2\m}}{t}
+ \frac{(\e + e^{-2 \m})^2}{2t^2}.
\]
Now
\beas
\n{j(t,r)}
&\leq&
\frac{\pi}{t^2}
\int_{-\infty}^\infty \int_0^\infty (1 + w^2 +F/t^2)^{1/4} f^{1/2}
\, \frac{\n{w}}{(1 + w^2 +F/t^2)^{1/4}} f^{1/2}  \,dF\,dw \\
&\leq&
\r (t,r)^{1/2} p(t,r)^{1/2}
\eeas
by the Cauchy-Schwarz inequality. Therefore
\[
\r^2 + p^2 - 2j^2 \geq \r^2 +p^2 -2 \r p = (\r -p)^2
\]
and
\beas
\frac{t^2}{8} K_2
&\geq&
\left( 4 \pi t (\r -p)\right)^2 - 4 \pi t (\r -p) \frac{\e +e^{-2\m}}{t}
+ \frac{(\e + e^{-2 \m})^2}{2t^2} \\
&=&
\left( 4 \pi t (\r -p) - \frac{\e + e^{-2 \m}}{2t} \right)^2
+ \frac{1}{4} \frac{(\e + e^{-2 \m})^2}{t^2} .
\eeas
Thus
\[
\left(R_{\alpha\beta\gamma\delta} R^{\alpha\beta\gamma\delta}\right) (t,r)
\geq 6\, \frac{\left(\e + e^{-2 \m (t,r)}\right)^2}{t^4} .
\]
Recalling (\ref{mudar}) we obtain
\[
e^{-2\m} +\e = \frac{e^{-2\mns} +\e}{t} - \frac{8 \pi}{t}
\int_1^t s^2 p(s,r)\, ds \geq  \frac{e^{-2\mns} +\e}{t} ,
\]
and inserting this into the previous estimate completes the proof.
\prfe
Next we show that the singularity at $t=0$ is also a crushing singularity
in the sense that the mean curvature of the surfaces of constant $t$
blows up as $t\to 0$.
\begin{thm} \label{crush}
Let $(f,\l,\m)$ be a solution of $(\ref{v})$--$(\ref{f4})$ on $]0,1]$
and define $c:= \inf e^{-\mns}$ for $\e=0$ or $\e=1$, and
$c:= \frac{3}{2} \inf (e^{-2\mns} -1)^{1/2}$ for $\e=-1$. Let
$k(t,r)$ denote the mean curvature of the surfaces of constant $t$.
Then
\[
k(t,r) \leq - c \, t^{-3/2} .
\]
\end{thm}
\prf
For a metric of the form
\[
ds^2 = - \phi^2 (t,x) dt^2 + g_{ij}dx^i dx^j
\]
where $i,j$ run from 1 to 3 the second fundamental form is given by
\[
k_{ij} = - (2 \phi)^{-1} \dot g_{ij},
\]
and its trace $k (t,x) = k_i^i (t,x)$ is the mean curvature of the surface
of constant $t$, at the point $x$. For the metric in our present situation
we obtain
\beas
k_{11} &=& - \frac{1}{2} e^{-\m} \dt (e^{2\l}) = - e^{2 \l -\m} \dot \l,\\
k_{22} &=& - t e^{-\m},\\
k_{33} &=& - t e^{-\m} \sin_\e \theta,
\eeas
and
\[
k(t,r) = - e^{-\m} \left(\dot \l + \frac{2}{t} \right),
\]
cf.\ (\ref{metric}). Recalling (\ref{lddar}) we obtain
\[
\dot\l = e^{2\mu} \left( 4 \pi t \r - \frac{\e +e^{-2\m}}{2t} \right)
\geq
-  e^{2\mu} \frac{\e +e^{-2\m}}{2t} ,
\]
and
\[
k(t,r) \leq e^\m \frac{\e - 3 e^{-2\m}}{2t}.
\]
If $\e =0$ or $\e=-1$ then
\[
k (t,r) \leq - \frac{3}{2t} e^{-\m},
\]
and the estimate
\[
e^{-2\m} \geq \frac{e^{-2\mns} +\e}{t}
\]
completes the proof of our assertion in these two cases. For $\e =1$
the estimate
\[
e^{-2\m} \geq \frac{e^{-2\mns}}{t} >1 =\e,
\]
which holds for $t$ small, shows that
\[
k(t,r) \leq - \frac{e^{-\m}}{t} \leq - \frac{e^{-\mns}}{t^{3/2}},
\]
and the proof is complete also in this case. \prfe
Since we have already determined the second fundamental form
it requires little additional effort to investigate
the quotients
\[
\frac{k^1_1}{k},\ \frac{k^2_2}{k},\ \frac{k^3_3}{k}
\]
in the limit $t\to 0$. The question whether these quotients
have limits is connected to the concept of a velocity dominated
singularity, cf.\ \cite{ELS}. It turns out that these limits exist and can be
determined in
the case of small initial data, more precisely:
\begin{thm} \label{veldo}
Let $(f,\l,\m)$ be a solution of $(\ref{v})$--$(\ref{f4})$ with small initial
data as described in Theorem~$\ref{glex}$. Then
\[
\lim_{t\to 0}\frac{k^1_1 (t,r)}{k(t,r)} = - \frac{1}{3},
\]
\[
\lim_{t\to 0}\frac{k^2_2 (t,r)}{k(t,r)} =
\lim_{t\to 0}\frac{k^3_3 (t,r)}{k(t,r)} =  \frac{2}{3},
\]
uniformly in $r \in \R$.
\end{thm}
\prf
We have
\beas
\frac{k^1_1 (t,r)}{k(t,r)}
&=&
\frac{t \dot \l (t,r)}{t \dot \l(t,r) + 2},\\
\frac{k^2_2 (t,r)}{k(t,r)}
&=& \frac{k^3_3 (t,r)}{k(t,r)} = \frac{1}{t \dot \l(t,r) + 2}.
\eeas
{}From (\ref{lddar}) we have
\[
t \dot \l = 4 \pi t^2 e^{2 \m} \r - \e \frac{e^{2\m}}{2} - \frac{1}{2} .
\]
As we have seen in the proof of Theorem~\ref{glex}
\[
e^{2 \m (t,r)} \leq C t,
\]
and the estimate on the $w$-support of $f$ in that theorem shows that
\[
\r (t,r) = \frac{\pi}{t^2} \int_{-P(t)}^{P(t)} \int_0^{F_0}
\vo f (t,r,w,F)\, dF\, dw \leq C t^{-3+c}
\]
so that
\[
4 \pi t^2 e^{2\m}\r (t,r) \leq C t^{c} \to 0,\ t \to 0.
\]
Thus
\[
t \dot \l (t,r) \to - \frac{1}{2},\ t \to 0,
\]
uniformly in $r$, and the proof is complete. \prfe

\noindent
{\bf Remark:}
As is easily checked a homogeneous and isotropic solution
in the case $\epsilon =0$ is given by
\[
f(t,w,F):=
\phi (t^2 w^2 + F),\
\l (t) :=
\mbox{ln}\, t,\
e^{-2\m (t)}=
\frac{8 \pi}{3} t^2 \r (t),
\]
where $\phi \in C^1 ([0,\infty[)$ is an arbitrary, nonnegative function with
compact support. For this solution, the limits considered in
Theorem~\ref{veldo}
are $1/3,1/3,1/3$ which shows that the result in that theorem need
not be true for solutions whose initial data violate the smallness
assumption.
\section{Going forward in time}
\setcounter{equation}{0}
Our main interest in this paper lies the behaviour of solutions
as $t$ approaches the singularity at $t=0$. Nevertheless, it is
possible to consider the solutions for $t\geq 1$. Our main result
here is that one can establish a bound on the $w$-support of $f$
and in spite of that the solutions need not exist for all $t\geq 1$.
Since this is in sharp contrast to the spherically symmetric,
asymptotically flat case, cf.\ \cite{RR}, and also to well known
results for the Vlasov-Poisson and Vlasov-Maxwell system we include
these considerations.

If one goes through the proof of Theorem~\ref{locex}, but
now for $t\geq 1$, one immediate problem is that
(\ref{mundar}) need not define $\m_n$ for all $t\geq 1$ but only
on some finite time interval. This results
in a more restrictive continuation criterion in the corresponding
local existence theorem the proof of which is omitted:
\begin{thm} \label{locexf}
Let $(\fn,\ln,\mn)$ be initial data as in Theorem~$\ref{locex}$.
Then there exists a unique, right maximal, regular solution $(f,\l,\m)$
of $(\ref{v})$--$(\ref{f4})$ with $(f,\l,\m)(1) = (\fn,\ln,\mn)$
on a time interval $[1,T[$ with $T \in ]1,\infty]$.
If
\[
\sup \Bigl\{ \n{w} \mid (t,r,w,F) \in \supp f \Bigr\} < \infty
\]
and
\[
\sup \Bigl\{ e^{2 \m(t,r)} \mid r \in \R,\ t \in [1,T[ \Bigr\} < \infty
\]
then $T=\infty$.
\end{thm}
The main result of this section is the following theorem:
\begin{thm} \label{wbound}
Assume that $(f,\l,\m)$ is a solution of $(\ref{v})$--$(\ref{f4})$ on a right
maximal interval of existence
$[1,T[$, and
\[
\sup \Bigl\{ e^{2 \m(t,r)} \mid r \in \R,\ t \in [1,T[ \Bigr\} < \infty .
\]
Then $T=\infty$.
\end{thm}
\prf
Assume that $T < \infty$. We show that under the assumption on $e^{2 \m}$
we obtain the bound
\[
\sup \Bigl\{ \n{w} \mid (t,r,w,F) \in \supp f \Bigr\} < \infty,
\]
which is a contradiction to Theorem~\ref{locexf}. The proof of the bound on
$w$ is similar in spirit to the proof of \cite[Theorem~2.1]{RRS}.
Define
\beas
P_+ (t)
&:=&
\sup \bigl\{ w \mid (r,w,F) \in \supp f(t) \bigr\},\\
P_- (t)
&:=&
\inf \bigl\{ w \mid (r,w,F) \in \supp f(t) \bigr\},
\eeas
assume that $P_+ (t) >0$ for some $t \in [1,T[$, and let $w(t) = w >0$
denote the $w$-component of a characteristic in $\supp f$.
We have
\beas
\dot w
&=&
\frac{4 \pi^2}{t} e^{2 \m} \int_{-\infty}^\infty \int_0^\infty
\left( \tilde w \vo - w \sqrt{1 + \tilde w^2 + \tilde F /t^2} \right)
f \, d\tilde F\, d\tilde w \\
&&
+ \frac{1 + \e e^{2 \m}}{2 t} w .
\eeas
Let us abbreviate
\[
\xi =  \tilde w \vo - w \sqrt{1 + \tilde w^2 + \tilde F /t^2}.
\]
As long as $w(s) >0$ we have the following estimates:
If $\tilde w \leq 0$ then $\xi \leq 0$. If $\tilde w > 0$ then
\beas
\xi
&=&
\frac{ \tilde w^2 (1 + w^2 + F/s^2) -
w^2 (1 + \tilde w^2 + \tilde F/s^2)}
{ \tilde w \vos + w \sqrt{1 + \tilde w^2 + \tilde F /s^2}}\\
&=&
\frac{ \tilde w^2 (1 + F/s^2) -
w^2 (1 + \tilde F/s^2)}
{ \tilde w \vos + w \sqrt{1 + \tilde w^2 + \tilde F /s^2}}\\
&\leq&
C \frac{\tilde w}{w(s)},
\eeas
and thus
\beas
\dot w (s)
&\leq&
C \frac{1}{s\,w(s)} \int_0^{\tilde P_+ (s)}\int_0^{F_0}
\tilde w f(s,r,\tilde w,\tilde F)\, d\tilde F\, d\tilde w +
\frac{C}{s} w(s)\\
&\leq&
\frac{C}{s} \left( \frac{\tilde P_+ (s)^2}{w(s)} + w(s) \right),
\eeas
where $\tilde P_+ := \max\{ P_+, 0 \}$.
Thus
\[
\frac{d}{ds} w^2 (s) \leq  \frac{C}{s} \tilde P_+ (s)^2
\]
as long as $w(s) >0$. Let $t_0 \in [1,t[$ be defined minimal such that
$w(s)>0$ for $s \in ]t_0,t]$. Then
\[
w^2(t) \leq w^2 (t_0) + C \int_{t_0}^t \frac{1}{s} \tilde P_+(s)^2 ds.
\]
Now either $t_0>1$ and $w(t_0)=0$ or $t_0=1$ and $w(t_0) \leq w_0$.
Thus
\[
\tilde P_+ (t)^2 \leq w_0^2 + C \int_{1}^t \frac{1}{s} \tilde P_+(s)^2 ds
\]
for all $t\in [1,T[$, since this estimate is trivial if $P_+ (t) \leq 0$.
If $T < \infty$ this estimate implies that $P_+$ is bounded on $[1,T[$.
Estimating $\dot w(s)$ from below in case $w(s)<0$ along the same lines
shows that $P_-$ is bounded as well, and the proof is complete.
\prfe
{\bf Remark:}
In the case of spherical symmetry it is easy to see that a solution cannot
exist for all $t\geq 1$, regardless of the size of initial data.
For such a solution $(f,\l,\m)$ the estimate
\beas
e^{-2 \m (t,r)}
&=&
\frac{e^{-\mns(r)} +1}{t} -1 - \frac{8 \pi}{t}
\int_1^t s^2 p(s,r)\, ds \\
&\leq&
\frac{e^{-\mns(r)} +1}{t} -1
\eeas
has to hold on the interval of existence $[1,T[$. Since the right hand side
of this estimate tends to $-1$ for $t\to \infty$ it follows that
$T < \infty$ and
\[
\nn{e^{2 \m (t)}} \to \infty,\ t \to T,
\]
by Theorem~\ref{wbound}. This should be compared with the behaviour of the
Schwarzschild metric
in the limit $r \nearrow 2 M$, i.\ e., approaching the event horizon from
inside of the black hole; cf.\ our interpretation of the metric
(\ref{metric}) in the introduction.


\begin{thebibliography}{10}

\bibitem{Chr}
D.~Christodoulou,
{\em Violation of cosmic censorship in the gravitational collapse of a
dust cloud},
Commun.\ Math.\ Phys.\
{\bf 93}, 171--195 (1984)

\bibitem{ELS}
D.~Eardley, E.~Liang, R.~Sachs,
{\em Velocity-dominated singularities in irrotational dust
cosmologies},
J.\ Math.\ Phys.\
{\bf 13}, 99--106 (1972)

\bibitem{Pf}
K.~Pfaffelmoser,
{\em Global classical solutions of the Vlasov-Poisson system in three
dimensions for general initial data},
J.\ Diff.\ Eqns.\
{\bf 95}, 281--303 (1992)


\bibitem{R}
G.~Rein,
{\em Static solutions of the spherically symmetric Vlasov-Einstein
system},
Math.\ Proc.\ of the Cambridge Phil.\ Soc.\
{\bf 115}, 559--570 (1994)


\bibitem{RR}
G.~Rein, A.~D.~Rendall,
{\em Global existence of solutions of the spherically symmetric
Vlasov-Einstein system with small initial data},
Commun.\ Math.\ Phys.\
{\bf 150}, 561--583 (1992)

\bibitem{RR3}
G.~Rein, A.~D.~Rendall,
{\em The Newtonian limit of the spherically symmetric Vlasov-Einstein system},
Commun.\ Math.\ Phys.\
{\bf 150}, 585--591 (1992)

\bibitem{RR4}
G.~Rein, A.~D.~Rendall,
{\em Smooth static solutions of the spherically symmetric Vlasov-Einstein
system},
Ann.\ de l'Inst.\ H.\ Poincar\'e, Physique Th\'eorique
{\bf 59}, 383--397 (1993)

\bibitem{RR2}
G.~Rein, A.~D.~Rendall,
{\em Global existence of classical solutions to the Vlasov-Poisson system
in a three-dimensional, cosmological setting},
Arch.\ Rational Mech.\ Anal.
{\bf 126}, 183--201 (1994)


\bibitem{RRS}
G.~Rein, A.~D.~Rendall, J.~Schaeffer,
{\em A regularity theorem for the spherically symmetric Vlasov-Einstein
system},
Commun.\ Math.\ Phys.,
to appear

\bibitem{Ren2}
A.~D.~Rendall,
{\em Global properties of locally spatially homogeneous cosmological
models with matter},
preprint 1994, gr-qc/9409009

\bibitem{Ren}
A.~D.~Rendall,
{\em Crushing singularities with spherical and plane symmetry},
in preparation

\bibitem{Sch}
J.~Schaeffer,
{\em Global existence of smooth solutions to the Vlasov-Poisson system
in three dimensions},
Commun.\ Part.\ Diff.\ Eqns.\
{\bf 16}, 1313--1335 (1991)


\end{thebibliography}
\end{document}